\documentclass[aps,prl,twocolumn,reprint,superscriptaddress,citeautoscript]{revtex4-2}

\usepackage{amsmath,amssymb,bm,mathtools}
\usepackage{physics}
\usepackage{hyperref}
\usepackage{graphicx}
\usepackage{xcolor}

\def\lesssim{\ \raise.3ex\hbox{$<$}\kern-0.8em\lower.7ex\hbox{$\sim$}\ }
\def\gesim{\ \raise.3ex\hbox{$>$}\kern-0.8em\lower.7ex\hbox{$\sim$}\ }

\begin{document}

\title{Thermal Spin Polarization Driven by Nuclear Spin-Orbit Coupling \\ in Neutron Star Pasta}

\author{Hiroyuki Tajima}
\email{hiroyuki.tajima@phys.s.u-tokyo.ac.jp}
\affiliation{Department of Physics, Graduate School of Science, The University of Tokyo, Tokyo 113-0033, Japan}
\affiliation{RIKEN Nishina Center, Wako 351-0198, Japan}
\affiliation{Quark Nuclear Science Institute, The University of Tokyo, Tokyo 113-0033, Japan}

\author{Yuta Sekino}
\email{sekino.yuta.y2@f.mail.nagoya-u.ac.jp}
\affiliation{Institute for Advanced Research, Nagoya University, Nagoya 464-8601, Japan}
\affiliation{Department of Physics, Nagoya University, Furo‑cho, Chikusa‑ku, Nagoya, Aichi 4648602, Japan}
\affiliation{Interdisciplinary Theoretical and Mathematical Sciences Program (iTHEMS), RIKEN, Wako, Saitama 351-0198, Japan}
\affiliation{Nonequilibrium Quantum Statistical Mechanics RIKEN Hakubi Research Team, RIKEN Cluster for Pioneering Research (CPR), Wako, Saitama 351-0198, Japan}

\author{Hiroshi Funaki}
\affiliation{Center for Spintronics Research Network, Keio University, Yokohama, 223-8522, Japan}
\affiliation{Kavli Institute for Theoretical Sciences, University of Chinese Academy of Sciences, Beijing, 100190, China.}

\author{Shota Kisaka}
\affiliation{Physics Program, Graduate School of Advanced Science and Engineering, Hiroshima University, Hiroshima 739-8526, Japan}
\affiliation{Hiroshima Astrophysical Science Center, Hiroshima University, 1-3-1, Kagamiyama, Higashi-Hiroshima, Hiroshima 739-8526, Japan}

\author{Nobutoshi Yasutake}
\affiliation{Department of Physics, Chiba Institute of Technology (CIT), 2-1-1 Shibazono, Narashino, Chiba 275-0023, Japan}
\affiliation{Advanced Science Research Center, Japan Atomic Energy Agency, Tokai, 319-1195, Japan}

\author{Mamoru Matsuo}
\email{mamoru@ucas.ac.cn}
\affiliation{Kavli Institute for Theoretical Sciences, University of Chinese Academy of Sciences, Beijing, 100190, China.}
\affiliation{CAS Center for Excellence in Topological Quantum Computation, University of Chinese Academy of Sciences, Beijing 100190, China}
\affiliation{Advanced Science Research Center, Japan Atomic Energy Agency, Tokai, 319-1195, Japan}
\affiliation{RIKEN Center for Emergent Matter Science (CEMS), Wako, Saitama 351-0198, Japan}

\date{\today}

\begin{abstract}
We discuss anomalous spin polarization on the surface of nuclear pasta in a neutron star,
driven by a nuclear spin-orbit interaction.
We present an effective two-band model of surface-localized neutrons near the nuclear pasta.
The central point is the emergence of a Rashba-type spin-orbit hybridization
generated by the neutron--nucleus spin-orbit force in the presence of
the strong density gradient normal to the pasta surface.
Starting from a single-particle Hamiltonian with a central potential and a standard nuclear spin-orbit interaction, we show the surface spin polarization occurs
due to the thermal inhomogeneity
even in the absence of a magnetic field.
Our study links neutron-star physics and solid-state spintronics and would contribute to understanding the interplay between spin dynamics and strong magnetic fields.
\end{abstract}

\maketitle

{\it Introduction}---
Neutron-star interiors host strongly nonuniform distributions of magnetic field, temperature, pressure, and composition, and an important challenge is to understand how such spatial structure is converted into angular-momentum dynamics in the inner crust~\cite{enoto2019observational}, manifested in pulsar glitches, vortex dynamics, mutual friction, and magnetic stress.
This issue becomes especially acute where ultrastrong magnetic fields, dense matter, and thermal nonequilibrium coexist on mesoscopic length scales.

Nuclear pasta, expected in the crust region, combines nontrivial geometry with strong modulation of the nuclear density~\cite{PhysRevLett.50.2066,hashimoto1984shape}.
An intriguing question is whether the pasta surface can convert local inhomogeneity into spin polarization, and thereby encode microscopic information about the crustal environment into angular-momentum dynamics in strongly magnetized neutron stars called magnetars~\cite{1992ApJ...392L...9D, 2017ARA&A..55..261K}.

From the viewpoint of nuclear structure, spin-orbit coupling is indispensable for the shell structure of finite nuclei and for the emergence of magic numbers~\cite{PhysRev.75.1969,PhysRev.75.1766.2}. Its dynamical role in neutron-star matter, however, remains largely unexplored. In homogeneous nuclear matter, the spin-orbit coupling is usually negligible; near nuclear surfaces, by contrast, it remains strong and persists even beyond the neutron drip line~\cite{PhysRevC.102.015802}. For surface neutrons confined near a slab interface of nuclear pasta, these conditions combine a surface-normal density gradient with a surviving spin-orbit field and thereby organize the spin dynamics at the pasta surface. 

\begin{figure}[t]
    \centering
    \includegraphics[width=1\linewidth]{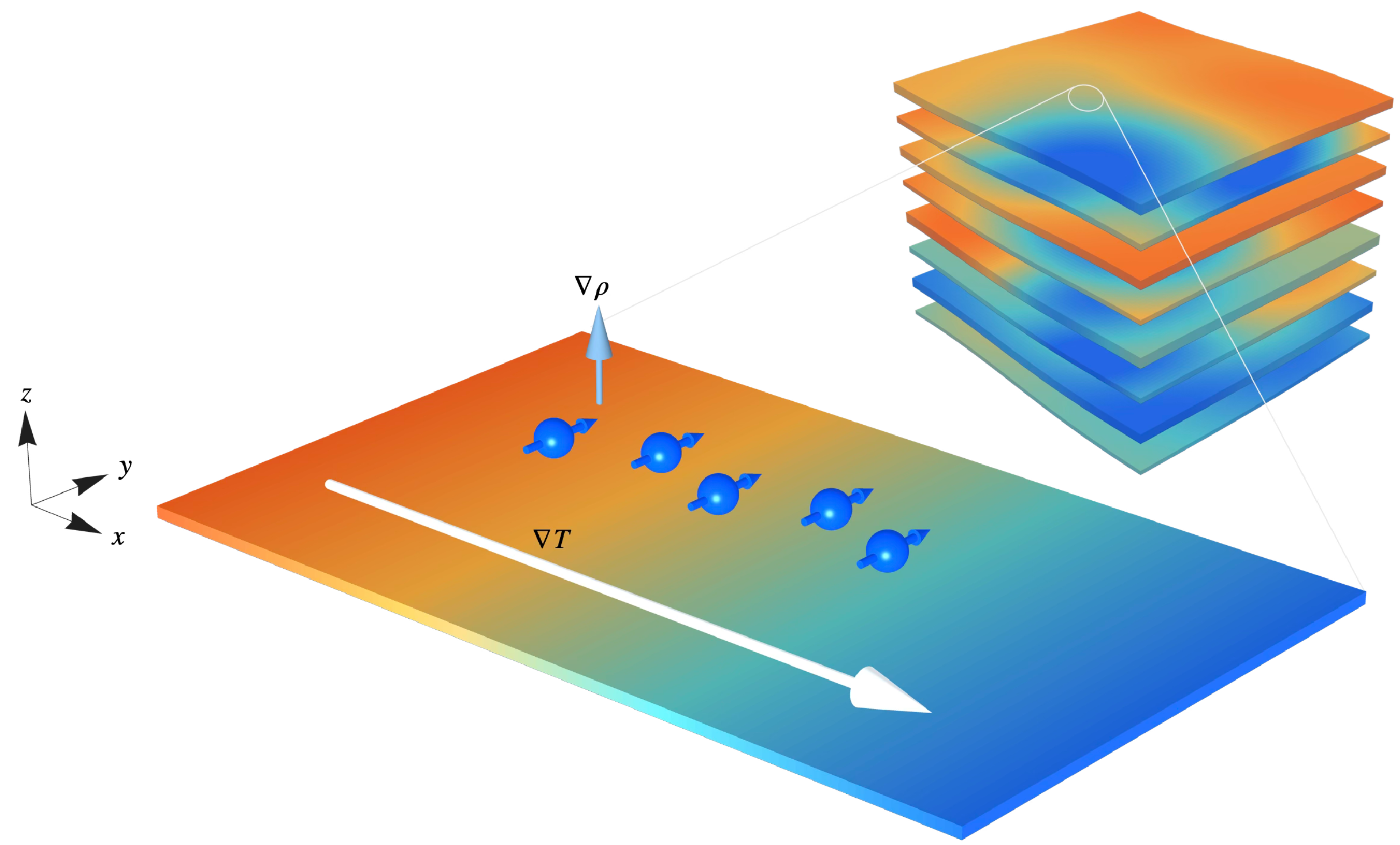}
    \caption{Schematic illustration of the thermal Rashba-Edelstein mechanism on a slab-pasta surface: the surface-normal density gradient $\nabla\rho$ and nuclear spin-orbit coupling generate a Rashba-like field for surface neutrons, and an in-plane thermal drive $\nabla T$ produces a surface spin polarization.}
    \label{fig:1}
\end{figure}

{A surface-normal density gradient of pasta breaks inversion symmetry, while dripped neutrons remain free to respond to an in-plane nonequilibrium drive, as at spin-orbit-coupled interfaces in condensed matter. At surfaces and interfaces with broken inversion symmetry, Rashba spin-orbit coupling generates spin-momentum locking and driven spin responses~\cite{bychkov1984oscillatory,Manchon2015}. In particular, the Rashba-Edelstein effect converts a nonequilibrium shift of Rashba-split states into a homogeneous spin density~\cite{Edelstein1990,Johansson2016}. This response has been explored in semiconductor heterostructures and at metallic or oxide interfaces~\cite{Yang2006,RojasSanchez2013,Lesne2016}, and is especially important for neutron-star matter, where a spin-current scenario may be a key to understanding its angular-momentum dynamics~\cite{funaki2025gyromagnetic,sedrakian2026spin}, but would require additional assumptions about transport, boundary conversion, and relaxation before one could infer a local accumulation.
It is also known that the spin-orbit coupling induces the parity mixing of Cooper pairs in superconductors without inversion symmetry~\cite{smidman2017superconductivity} and the Edelstein effect remains therein~\cite{PhysRevLett.75.2004}.
While this suggests that Rashba-type spin–orbit coupling modifies neutron superfluidity, the Edelstein response in the normal state itself is already worth investigating due to the relevance to the thermal pasta phase above the superfluid critical temperature.

In this work, we formulate an effective low-energy theory for neutrons near the surface of slab nuclear pasta and show that the combination of the surface density gradient and the nuclear spin-orbit coupling produces a Rashba-type hybridization between spins of surface-localized neutrons. Considering the linear response to an in-plane thermal drive, we show that the surface neutrons become spin-polarized through a thermal Rashba-Edelstein effect even in the absence of a magnetic field. 

Since the surface magnetic fields of magnetars are on the order of $10^{14}–10^{15} \mathrm{G}$~\cite{enoto2019observational}, it is not unreasonable to expect that the magnetic field in the inner crust may locally reach $10^{16}\,\mathrm{G}$. Therefore, in this paper, we assume magnetic fields up to $10^{17}\,\mathrm{G}$ in order to explore their effects. Such strong fields are introduced simply to examine magnetic effects; the key point to emphasize is that the spin polarization of interest in this work can arise even in the absence of a magnetic field.

{\it Microscopic model for neutrons near the nuclear pasta surface}---
We start from the second-quantized form of the effective Hamiltonian, using units with $\hbar=k_{\rm B}=1$,
\begin{align}
    \hat{H}
    =
    \int d^3\bm{r}\,
    \hat{\Psi}^\dagger(\bm{r})
    \hat{h}(\bm{r})
    \hat{\Psi}(\bm{r}),
\end{align}
where the neutron field operator $\hat{\psi}_{\sigma}(\bm{r})$ with spin $\sigma=\uparrow,\downarrow$ is written in a spinor form as $\hat{\Psi}(\bm{r})=(        \hat{\psi}_\uparrow(\bm{r}),
        \hat{\psi}_\downarrow(\bm{r}))^{\rm T}$.
The Hamiltonian density operator $\hat{h}(\bm{r})$
reads
\begin{align}
    \hat{h}(\bm{r})
    =
    \frac{\hat{\bm{p}}^2}{2m^*}
    +U(\bm{r})
+\hat{V}_{\rm SO}(\bm{r})
    -\mu_n \bm{B}\cdot \bm{\sigma},
\end{align}
where
$\hat{\bm{p}}=-i\nabla$ is the momentum operator,
$m^*$ is the neutron effective mass, $U(\bm{r})$ is the mean-field potential, $\hat{V}_{\rm SO}(\bm{r})$ is the spin-orbit term,
$-\mu_n\bm{B}\cdot\bm{\sigma}$ is the Zeeman energy induced by the magnetic field $\bm{B}$ with neutron magnetic moment $\mu_n$ and the Pauli matrix $\bm{\sigma}=(\sigma_x,\sigma_y,\sigma_z)$. 
In terms of the Skyrme-type Hartree-Fock theory~\cite{PhysRevC.5.626,ring1983nuclear,RevModPhys.75.121}, the standard one-body
spin-orbit potential is given by
\begin{align}
    \hat{V}_{\rm SO}(\bm{r})=\bm{W}_{n}(\bm{r})\cdot(\hat{\bm{p}}\times\bm{\sigma}),
\end{align}
where 
   $ \bm{W}_n(\bm{r})=W_0\bm{\nabla}\rho_n(\bm{r})$ 
is the local spin-orbit strength with the neutron density $\rho_n(\bm{r})$.
Here, we consider the nuclear-pasta surface where the density gradient is given by $\partial_z\rho_n(z)$ and neglect those in the $x$ and $y$ directions.
Accordingly, we obtain 
\begin{align}
    \hat{V}_{\rm SO}(\bm{r})=W_0(\partial_z\rho_n)(\hat{p}_x\sigma_y-\hat{p}_y\sigma_x),
\end{align}
which is known as a Rashba spin-orbit coupling in the context of condensed-matter physics~\cite{rashba1960properties,bychkov1984oscillatory}.

To formulate a low-energy theory applicable to various pasta geometries, we decompose the single-particle Hilbert space into the surface sector and its complement as $\mathcal{H} = \mathcal{H}_S \oplus \mathcal{H}_B$,
where $\mathcal{H}_S$ denotes the subspace of modes localized near the nuclear-pasta surface, while $\mathcal{H}_B$ denotes the complementary extended sector. Accordingly, the field operator is expanded as
\begin{align}
\label{eq:5}
    \hat{\psi}_\sigma(\bm{r})
    =
    \sum_{\alpha\in \mathcal{H}_S }
    \phi_{\alpha\sigma}(\bm{r}) \hat{c}_{\alpha\sigma}
    +
    \sum_{\beta\in \mathcal{H}_B}
    \chi_{\beta\sigma}(\bm{r})\, \hat{d}_{\beta\sigma},
\end{align}
where $\phi_{\ell\sigma}(\bm{r})$ represents a surface-supported mode with set of quantum numbers $\alpha$ and $\chi_{\bm{k}\sigma}(\bm{r})$ represents an extended mode with set of quantum numbers $\beta$
in the complementary sector.
$\hat{c}_{\alpha\sigma}$ and $\hat{d}_{\beta\sigma}$ are annihilation operators of surface-supported and extended modes, respectively.
With this decomposition, the Hamiltonian can be written as $ \hat{H} = \hat{H}_S + \hat{H}_B + \hat{H}_{SB}$,
where
\begin{align}
    \hat{H}_S
    &=
    \sum_{\alpha,\alpha' \in \mathcal{H}_S}
    \sum_{\sigma}
    \hat{c}_{\alpha\sigma}^\dagger
    \mel{\phi_{\alpha\sigma}}{\hat{h}}{\phi_{\alpha'\sigma}}
    \hat{c}_{\alpha'\sigma}, \\
    \hat{H}_B
    &=
    \sum_{\beta,\beta' \in \mathcal{H}_B}
    \sum_{\sigma}
    \hat{d}_{\beta\sigma}^\dagger
    \mel{\chi_{\beta\sigma}}{\hat{h}}{\chi_{\beta'\sigma}}
    \hat{d}_{\beta'\sigma}, \\
    \hat{H}_{SB}
    &=
    \sum_{\alpha \in \mathcal{H}_S,\beta \in \mathcal{H}_B}
    \sum_{\sigma}
    \left[
        \hat{c}_{\alpha\sigma}^\dagger
        \mel{\phi_{\alpha\sigma}}{\hat{h}}{\chi_{\beta\sigma}}
        \hat{d}_{\beta\sigma}
        + \mathrm{h.c.}
    \right].
\end{align}

Practically, we use the single-mode approximation for the surface mode in the $z$ direction and the plane waves in $x,y$ directions as $\phi_{\alpha\sigma}(\bm{r})\simeq \frac{1}{\sqrt{A}}e^{i\bm{k}_\perp\cdot\bm{r}_\perp}\tilde\phi(z)$ where $\sqrt{A}$ is the area of the nuclear-pasta surface and $\bm{k}_\perp=(k_x,k_y)$ is the wavevector perpendicular to the $z$-axis.
The single-localized mode can be described by
\begin{align}
    \tilde\phi(z)=\sqrt{\frac{2}{\xi}}e^{-(z-R)/\xi}\theta(z-R),
\end{align}
where $\xi$ is the localization length and $R$ is the nuclear size.
The extended mode is assumed to be a plane wave in free space as $\chi_{\beta\sigma}(\bm{r})\simeq\frac{1}{V}e^{i\bm{k}\cdot\bm{r}}$ where $V$ is the volume of the bulk space.
Eventually, Eq.~\eqref{eq:5} reduces to
\begin{align}
    \hat{\psi}_{\sigma}(\bm{r})
    \simeq \frac{1}{\sqrt{A}}\sum_{\bm{k}_\perp}
    e^{i\bm{k}_\perp\cdot\bm{r}_\perp}\tilde{\phi}(z)\hat{c}_{\bm{k}_\perp}
    +\frac{1}{\sqrt{V}}\sum_{\bm{k}}e^{i\bm{k}\cdot\bm{r}}\hat{d}_{\bm{k}\sigma}.
\end{align}
Then, the kinetic Hamiltonian is rewritten as
\begin{align}
   \hat{H}_{S}
    =
    \sum_{\bm k_\perp}
    \hat{C}_{\bm k_\perp}^\dagger
    \left[
    \varepsilon_{\bm{k}_\perp}\sigma_0
    +\bm{d}({\bm{k}_\perp})\cdot\bm{\sigma}
    \right]
    \hat{C}_{\bm k_\perp},
\end{align}
and
\begin{align}
    \hat{H}_{B}
    =
    \sum_{\bm{k}}
    \hat{D}_{\bm k}^\dagger
    \left[
    (\varepsilon_{\bm{k},b}+\Delta\epsilon)\sigma_0
    -\mu_n\bm B\cdot\bm\sigma
    \right]
    \hat{D}_{\bm k},
\end{align}
with $\hat{C}_{\bm{k}_\perp}=(\hat{c}_{\bm{k}_\perp\uparrow},\,\hat{c}_{\bm{k}_\perp\downarrow})^{\rm T}$, $\hat{D}_{\bm{k}}=(\hat{d}_{\bm{k}\uparrow},\,\hat{d}_{\bm{k}\downarrow})^{\rm T}$, $\sigma_0=\mathrm{diag}(1,1)$, 
and 
\begin{align}
\label{eq:dvector1}
\bm{d}({\bm k_\perp})
    =
    \left(
        -\alpha_s k_y-\mu_n B_x,\,
        \alpha_s k_x-\mu_n B_y,\,
        -\mu_n B_z
    \right),    
\end{align}
where $\varepsilon_{\bm{k}_\perp}={k_\perp^2}/{2m^*}$ and $\varepsilon_{\bm{k},b}=k^2/2m^*+\Delta \epsilon$
are the kinetic energies of surface and bulk neutrons, respectively.
In Eq.~\eqref{eq:dvector1}, $\alpha_s=W_0\partial_z\rho_z$ is the effective Rashba spin-orbit coupling strength.
For bound surface neutrons, the level difference $\Delta\epsilon$ is positive and equivalent to the binding energy $E_{\rm b}$.
For the resonant case, we may have $\Delta\epsilon<0$ but the hybridization term $\hat{H}_{SB}$ induces the imaginary part, i.e., $\Delta\epsilon\in\mathbb{C}$, due to the decay into the bulk mode.
Even in such a case, the localization length can be defined as $\xi\simeq {\rm Re}\left[1/\sqrt{2m\Delta\epsilon}\right]$~\cite{tajima2026non}.
At low energies, the coupling to the extended bulk modes can be absorbed into a renormalization of the surface parameters.
We therefore focus on $\hat{H}_S$.
The corresponding band dispersions are given by
\begin{align}
    E_{\kappa=\pm}(\bm k_\perp)
    =
    \varepsilon_{\bm k_\perp}+\kappa |\bm d(\bm k_\perp)|,
\end{align}
which is plotted in Fig.~\ref{fig:2} as a function of $k_\perp=|\bm{k}_\perp|$ under  $\alpha_s = -2\,{\rm MeV\cdot fm}$ (see also the estimation of $\alpha_s$ in the Appendix).
One can see the splitting of the surface dispersion due to the spin-orbit coupling.
For $B_z=0$, the bottom of the band is shifted to $k_\perp=m\alpha_s$.
In the case of a strong magnetic field,
the band-crossing point at $k_\perp$ becomes gapped, and its energy gap is given by $2|\mu_nB_z|$.
However, we note that the dispersion is not strongly affected by smaller magnetic fields (e.g., $10^{12-15}\,{\rm G}$).

\begin{figure}[t]
    \centering
    \includegraphics[width=1\linewidth]{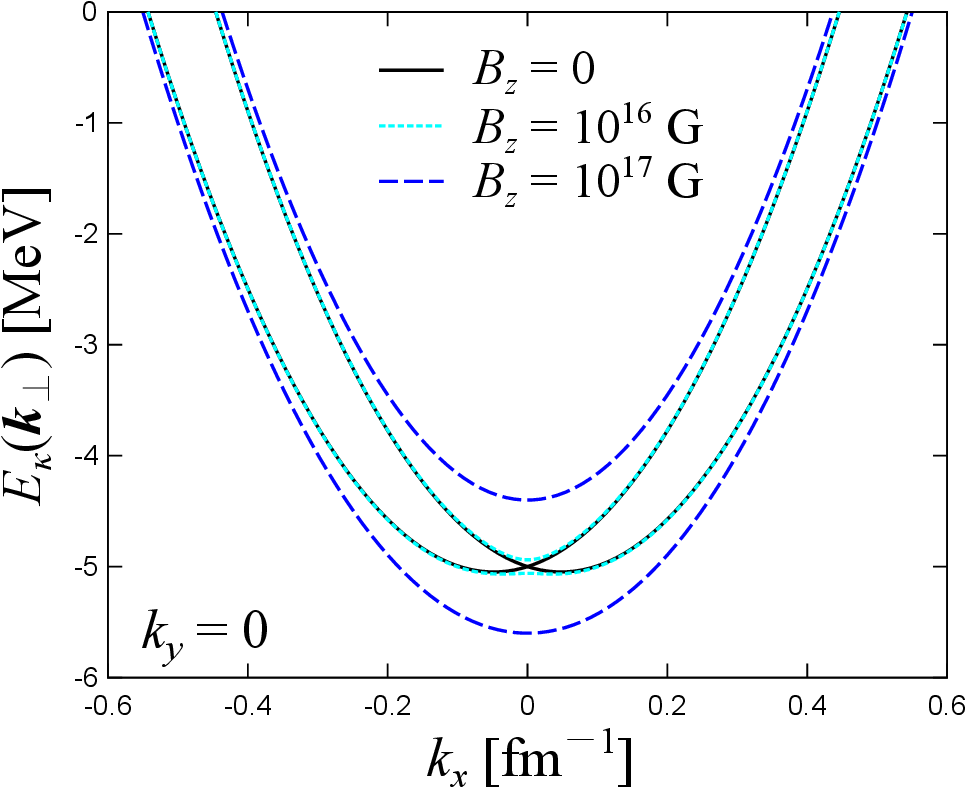}
    \caption{
     Dispersion relation $E_{\kappa}(\bm{k}_\perp)$ of the surface neutrons
     measured from the continuum bottom of the extended modes, where we set $\Delta\epsilon=5\,{\rm MeV}$,
     at $\alpha_s=-2\,{\rm MeV}\cdot{\rm fm}$,
     and $k_y=0$. 
     The magnetic fields are taken as $B_z=0,\,10^{16}\,{\rm G},\,10^{17}\,{\rm G}$.
     }
    \label{fig:2}
\end{figure}

{\it Rashba-Edelstein effect}---
We now make the connection to a spin polarization of surface neutrons due to the Rashba-Edelstein effect.
For the present Rashba-split band in the inset of Fig.~\ref{fig:2}, the expectation value of the Pauli spin operator $\langle \bm{\sigma}\rangle_{\kappa,\bm{k}_\perp}$
is related to the direction of the $\bm d$ vector as
\begin{align}
\langle{\bm \sigma}\rangle_{\kappa,\bm{k}_\perp}
=
\kappa\frac{\bm d(\bm{k}_\perp)}{|\bm d(\bm{k}_\perp)|}.
\end{align}
In this regard, the two bands $\kappa=\pm$ carry opposite spin textures in momentum space as $\langle{\bm \sigma}\rangle_{+,\bm{k}_\perp}=-\langle{\bm \sigma}\rangle_{-,\bm{k}_\perp}$.
Then, the spin expectation value reads $s_\kappa^j(\bm{k}_\perp)
=
\frac{1}{2}\,
\ev{\sigma_j}_{\kappa,\bm k}$.
Using the semiclassical description of the nonequilibrium distribution, we obtain the homogeneous spin polarization density
\begin{align}
    \delta S_j=\sum_{\kappa=\pm}\int\frac{d^2\bm{k}_\perp}{(2\pi)^2}s_\kappa^j(\bm{k})\delta f_\kappa(\bm{k}_\perp)
\end{align}
where 
\begin{align}
    \delta f_\kappa(\bm{k}_\perp)=-\tau[\nabla_\perp T\cdot\nabla_{\bm{k}_\perp}E_{\kappa}(\bm{k}_\perp)]\frac{\partial f(E_{\kappa}(\bm{k}_\perp))}{\partial T}
\end{align}
is the deviation of the distribution function
induced by the in-plane linear temperature gradient $\nabla_\perp T$
with the in-plane momentum relaxation time $\tau$ and the equilibrium Fermi distribution function $f(x)=1/(e^{(x-\mu_\perp)/T}+1)$.
$\mu_\perp$ is the chemical potential conjugate with the in-plane density $\rho_\perp$ and taken to be equal to $\Delta\epsilon$ such that the surface band is fully occupied.
The Edelstein susceptibility $\chi_{ij}$ is defined as
\begin{align}
    \delta S_i=\chi_{ij}(\nabla_\perp T)_j.
\end{align}

When the homogeneous magnetic field is applied normal to the nuclear-pasta surface, $\bm B=(0,0,B_z)$,
we obtain
   $ E_\kappa(\bm k_\perp)
    =
    \varepsilon_{\bm k_\perp}
    +\kappa
    \sqrt{
        \alpha_s^2 k_\perp^2
        +
        (\mu_n B_z)^2
    }$.
Accordingly, we obtain
\begin{align}
    \chi_{xy}&= \frac{\alpha_s\tau}{8\pi m^*}\sum_{\kappa=\pm}
    \int_0^{\infty} 
    \frac{k_\perp^3dk_\perp}{|\bm{d}(\bm{k}_\perp)|}
    \left(
    {\kappa}+\frac{m^*\alpha_s^2}{|\bm{d}(\bm{k}_\perp)|}
    \right)\cr
    &\quad\quad\quad\times
    \frac{\partial f(E_\kappa(\bm{k}_\perp))}{\partial T},
\end{align}
and $\chi_{yx}=-\chi_{xy}$.

\begin{figure}[t]
    \centering
    \includegraphics[width=1\linewidth]{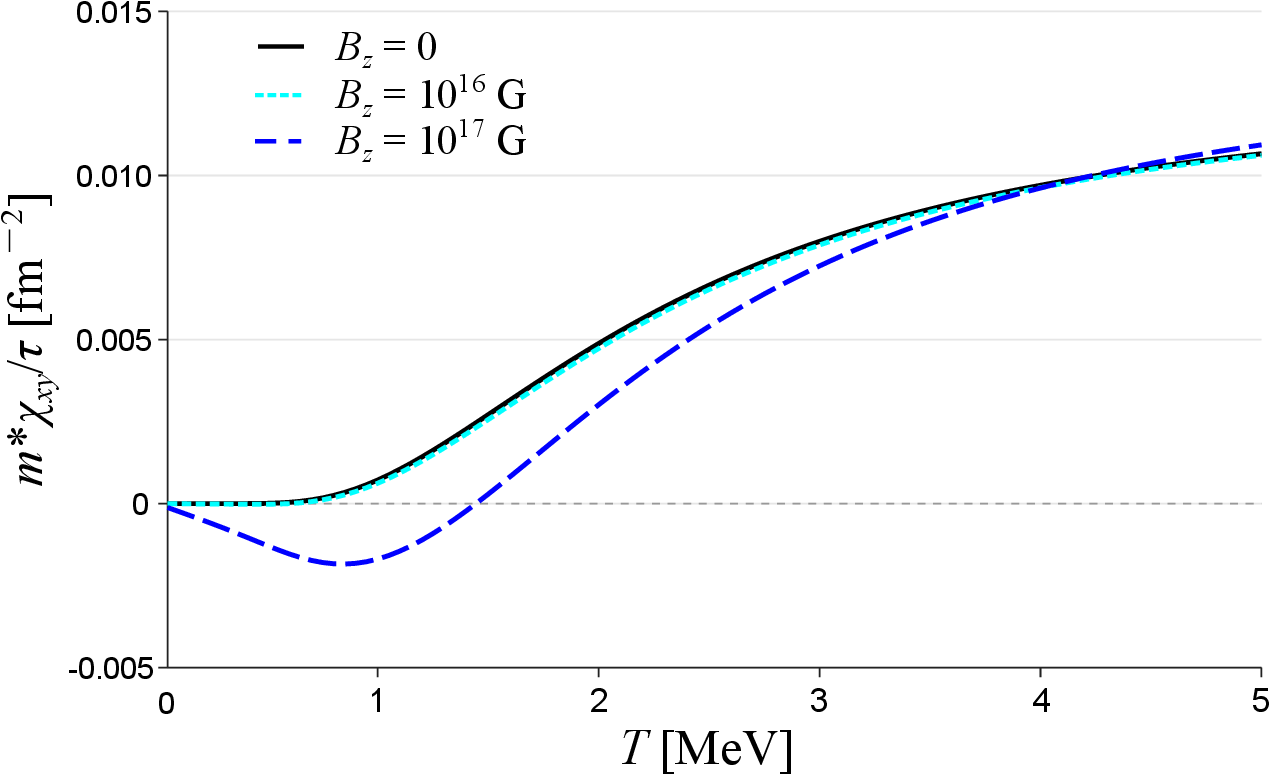}
    \caption{Normalized Edelstein susceptibility $m^*\chi_{xy}/\tau$ at the surface of nuclear pasta as a function of the temperature $T$, where $\alpha_s=-2\,{\rm MeV\cdot fm}$ and $\mu_\perp=5\,{\rm MeV}$. The magnetic field is taken as $B_z=0,\, 10^{16}\,{\rm G},\, 10^{17}\,{\rm G}$.}
    \label{fig:3}
\end{figure}

Figure~\ref{fig:3} shows the normalized Edelstein susceptibility $m^*\chi_{xy}/\tau$ as a function of $T$.
It is found that
$\chi_{xy}$ is nonzero at finite temperatures, indicating the spontaneous in-plane spin polarization due to the thermal bias.
Importantly, the spin polarization induced by the Rashba-Edelstein effect occurs even with vanishing or weak magnetic fields (e.g., $B_z\sim 10^{12-15}\,\mathrm{G}$).
While $\chi_{xy}$ increases with increasing $T$,
$\chi_{xy}$ is nonzero even above the typical superfluid critical temperature$\sim 1 \,{\rm MeV}$~\cite{sedrakian2019superfluidity}.
The non-monotonic behavior of $m^*\chi_{xy}/\tau$ at $B_z=10^{17}\,{\rm G}$ is associated with the Zeeman energy $2|\mu_nB_z|\simeq 1.2\,{\rm MeV}$.
Although the $T$-dependence of $\chi_{xy}$ changes at different $\mu_\perp$, Fig.~\ref{fig:3} indicates that the order of $\chi_{xy}$ remains unchanged in the typical temperature range of neutron stars with stable pasta phases~\cite{PhysRevC.77.035806,PhysRevC.87.055805,PhysRevD.106.063020}. 

{\it Discussion}---
As $\chi_{xy}$ is proportional to $\tau$, the spin polarization through the Rashba-Edelstein effect depends on the in-plane momentum relaxation, which is governed by decays toward bulk-extended mode, fluctuations or disorders in the pasta, as well as the neutron-neutron interaction. In this sense, we may write
\begin{align}
    \frac{1}{\tau}=\nu_{\rm dis}+\nu_{nn}+\nu_{\rm bulk},
\end{align}
where 
$\nu_{\rm dis}$ is the in-plane collision rate with disorders in the nuclear pasta,
$\nu_{nn}$ is the neutron-neutron collision rate,
and $\nu_{\rm bulk}$ is the decay rate toward the bulk-extended mode.
Based on the relativistic mean-field study~\cite{PhysRevD.106.063020}, the disorder effect is important at high temperatures ($T\gesim 5\, {\rm MeV}$, but depending on the proton fraction) and hence $\nu_{\rm dis}$ is negligible at low temperatures.
Considering the analogy with unitary Fermi liquid realized in ultracold atoms~\cite{PhysRevLett.122.093401,PhysRevA.99.063606},
$\nu_{nn}$ may be proportional to $T^2$ at low temperatures and decreases with $T^{-1/2}$ in the Boltzmann regime.
In this sense, at low temperatures, $\tau$ is expected to be dominated by $\nu_{\rm bulk}$, which is associated with the self-energy for the surface-bulk hybridization (see also Appendix) as $\nu_{\rm bulk}\simeq -2{\rm Im}\Sigma_{\sigma}(\bm{k}_\perp,\omega)$.
Since our effective model is valid when $|{\rm Im}\Sigma_\sigma(\bm{k}_\perp,\omega)|\ll\Delta\epsilon$, we have the lower bound of $\tau$, that is, $\tau\gg 1/|\Delta\epsilon|\sim O(1)\, {\rm MeV}^{-1}$.

Our scenario may contribute to the spin polarization in the neutron-star crusts,
which assists the spin-triplet pairing~\cite{PhysRevC.108.L052802,yoshimura2026superfluid}, and also bridges the study for the possible spin-triplet pairing in heavy nuclei with spin-orbit coupling~\cite{PhysRevC.109.034302}.
Since we consider the linear-response regime near the unpolarized system, our prediction does not apply to the large spin-polarization regime.
Nevertheless, we can estimate the condition of $\tau\nabla_\perp T$ where the Edelstein-induced spin polarization $P= |\delta S_x|/\rho_{\perp}$ is non-negligible as $P=O(10^{-2})$, noting that $\rho_\perp=\sum_{\kappa}\int\frac{d^2\bm{k}_\perp}{(2\pi)^2}f(E_\kappa(\bm{k}_\perp))=O(10^{-2})\, {\rm fm}^{-2}$ in the present setup.
Based on the result in Fig.~\ref{fig:3},
the estimated $|\delta S|$ is given by
\begin{align}
    |\delta S|\, 
    &\sim 
     O(10^{-3}) \,{\rm fm}^{-1}\times\tau{|\nabla_\perp T|},
\end{align}
indicating that $P=O(10^{-2})$ is achieved when
\begin{align}
\label{eq:22}
    \tau|\nabla_\perp T|=O(0.1)\,{\rm fm}^{-1}.
\end{align}
If we assume $\tau=O(1)\,{\rm MeV}^{-1}$,
Eq.~\eqref{eq:22} indicates $|\nabla_\perp T|=O(0.1)\,{\rm MeV/fm}$ gives a non-negligible spin polarization. 
Possible sources of $\nabla_\perp T$ include anisotropic heat transport in nuclear pasta and transient thermal inhomogeneities during cooling. In particular, when the system crosses the neutron superfluid transition, nonequilibrium order-parameter textures and vortices may be generated through the Kibble-Zurek mechanism~\cite{ko2019kibble,10.1093/mnras/stae1642}. Although the Kibble-Zurek mechanism does not directly determine $\nabla_\perp T$, the resulting vortices and textures can affect both the local heat flow and the momentum relaxation time entering the Edelstein response.

In the slab phase we studied in this work, the periodic structure of pasta nuclei is expected~\cite{PhysRevC.105.045807}.
If the nearly symmetric nuclear surfaces face each other,
the surface spin-orbit coupling might be weakened due to the interference between opposite spin textures~\cite{PhysRevB.105.205429}.
On the other hand, once if there is an asymmetry between two nuclear surfaces, the Rashba-Edelstein effect could be induced by the spin-orbit coupling.
In this sense, more complicated geometries of the nuclear pasta may induce the large spin polarizations, which could be an intriguing seed of gyromagnetic angular momentum interconversion~\cite{funaki2025gyromagnetic}.

While we discuss the out-of-plane magnetic field effect on the Edelstein susceptibility, the in-plane magnetic field does not change our main conclusion.
However, if the spin-polarization directions induced by the in-plane Zeeman effect and the Rashba-Edelstein effect
are opposite each other, the resulting spin polarizations may partially cancel each other out.

{\it Summary}---
In this work,
we have theoretically discussed the role of spin-orbit force on the nuclear-pasta surface in a neutron-star crust.
The density gradient of nuclear pasta induces the Rashba-type spin-orbit coupling on the surface, which has been studied widely in the spintronics community.
Using the effective two-band model for surface-localized neutrons on the pasta nuclei,
we have shown that neutrons exhibit the Rashba-Edelstein effect, where the transverse spin polarization is induced by thermodynamic drives such as temperature inhomogeneity.
The local interplay between spin and thermal interconversion studied in this work may lead to  large-scale dynamics of a neutron star, which will be left for future work.

\begin{acknowledgments}
H. T. is grateful to the Nuclear Theory group member at The University of Tokyo for the valuable discussions during the early stages of this work.  
We acknowledge JSPS KAKENHI for Grants 
(Nos.~JP21K03436, 
 JP22K13981,
 JP23H01839,
 JP22K03681,
 JP23K22429,
 JP23K22538,
 JP23K20841,
 JP24K07054,
 JP24H00322,
 JP25K17351,
 and JP26K07063).
Y.~S. is supported by the RIKEN TRIP initiative (RIKEN Quantum).
M. M. is supported by the National Natural Science Foundation of China (NSFC) under Grant No. 12374126 and
by the Priority Program of Chinese Academy of Sciences under Grant No. XDB28000000.
\end{acknowledgments}

\bibliographystyle{apsrev4-1}
\bibliography{reference.bib}

\appendix
\section{Hybridization term}
The hybridization term between surface-localized and extended-bulk modes is taken to be spin independent and expressed as
\begin{align}
    H_{SB}
    =
    \sum_{\bm k_\perp,k_z}
    \left[
        C_{\bm k_\perp}^\dagger \Lambda_{\bm{k}}\sigma_0 D_{\bm k}
        +
        D_{\bm k}^\dagger \Lambda_{\bm{k}}^*\sigma_0 C_{\bm k_\perp}
    \right],
\end{align}
where $\Lambda_{\bm{k}}=\langle\phi_{\bm{k}_{\perp\sigma}}|\hat{h}|\chi_{\bm{k}\sigma}\rangle$ is the transition matrix element.

Within the second-order perturbation,
we obtain the self-energy
\begin{align}
    \Sigma_{\sigma}(\omega,\bm k_\perp)
    =
    \sum_{k_z}
    \frac{|\Lambda_{\bm{k}}|^2}
        {\omega+i0^+-
        \varepsilon_{\bm{k},b}-\Delta\epsilon
        +\mu_n{B_z}{\sigma}
        }.
\end{align}
Generally, the surface band is well-defined when
$|{\rm Im}\Sigma_\sigma(\omega,\bm{k}_\perp)|\ll |\Delta\epsilon|$, which can be regarded as a criterion of our effective model.

\section{Estimation of Rashba spin-orbit coupling strength }
Here we estimate the effective spin-orbit coupling strength 
\begin{align}
    \alpha_s=\int_0^\infty dz\,|\tilde{\phi}(z)|^2\alpha(z).
\end{align}
For simplicity, we consider the neutron-dominated case
\begin{align}
    \alpha(z)=\frac{W_0}{2}\partial_z[2\rho_n(z)+\rho_p(z)]
    \simeq W_0\partial_z\rho_n(z),
\end{align}
where $\rho_{n(p)}$ is the neutron (proton) density, 
and assume the Woods-Saxon-like density profile
\begin{align}
    \rho_n(z)=\frac{\rho_{n,0}}{1+e^{(z-R)/a}},
\end{align}
where $a$ is the diffuseness parameter and $\rho_{n,0}$ is the bulk neutron density.
The explicit form of $\alpha_s$ reads
\begin{align}
    \alpha_s
    &=-\frac{2}{\xi a}
    W_0\rho_{n,0}
    \int_0^{\infty}
    dz\,
    e^{-\frac{2z}{\xi}}\frac{e^{z/a}}{(1+e^{z/a})^2}.
\end{align}
It is convenient to introduce $x=e^{-z/a}$ as
\begin{align}
    \alpha_s&=-\frac{2}{\xi }W_0\rho_{n,0}\int_0^{1}
    dx\frac{x^{2a/\xi}}{(1+x)^2}\cr
    &\equiv-\frac{W_0\rho_{n,0}}{\xi}I(a/\xi),
\end{align}
where the integral $I(a/\xi)$ is plotted in Fig.~\ref{fig:4}.
Note that for $a\ll\xi$, we get
\begin{align}
    \alpha_s=-\frac{W_0\rho_{n,0}}{\xi}+O(a/\xi).
\end{align}
Using $W_0=123\,{\rm MeV}\cdot{\rm fm}^5$ and $\rho_{n,0}=0.08\,{\rm fm}^{-3}$, and $\xi=3\sim 8 \,{\rm fm}$ (which is larger than $a\sim 1\,{\rm fm}$~\cite{PhysRevC.102.015802}),
we obtain $|\alpha_s|\simeq 1.2\sim3.3\,{\rm MeV}\cdot{\rm fm}$.
Together with a typical momentum $k_\perp\simeq 1\,{\rm fm}^{-1}$,
the spin-orbit energy may be given by $E_{\rm so}\sim\alpha_s k_\perp\simeq 1\,{\rm MeV}$,
which is close to the estimation of the local spin-orbit coupling strength in Ref.~\cite{PhysRevC.102.015802}.

\begin{figure}[t]
    \centering
    \includegraphics[width=0.9\linewidth]{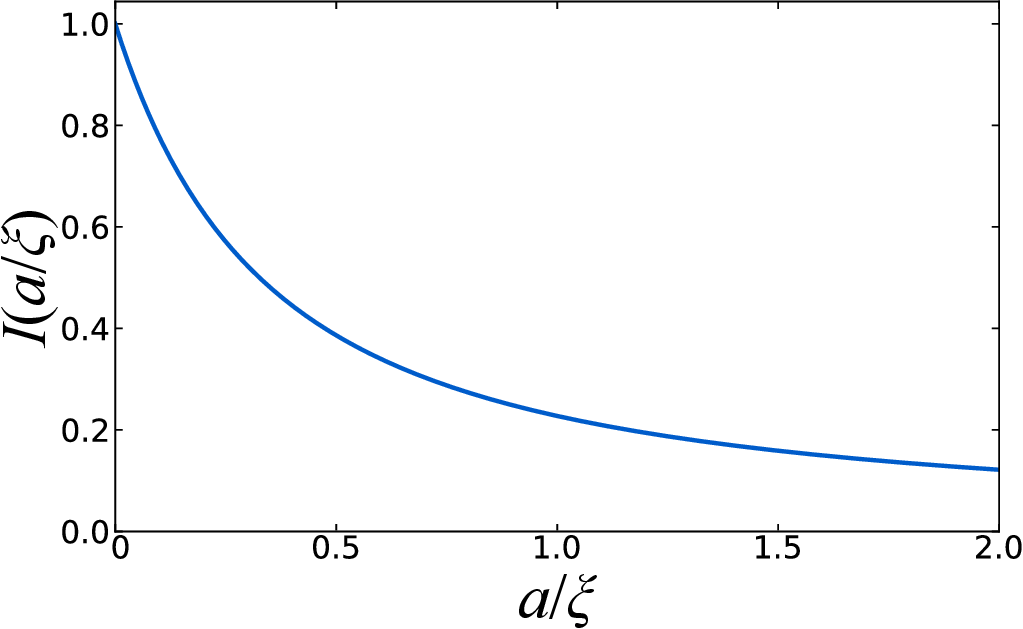}
    \caption{Integral $I(a/\xi)=2\int_0^1dx\frac{x^{2a/\xi}}{(1+x)^2}$ for the evalution of the Rashba-type spin-orbit coupling strength $\alpha_s$.}
    \label{fig:4}
\end{figure}

\section{Derivation of Edelstein susceptibility}
We derive the expression of $\chi_{xy}$ shown in the main text.
For $\bm{B}=(0,0,B_z)$,
we obtain
\begin{align}
    \partial_{k_x}E_\kappa(\bm{k}_\perp)
    &=\frac{k_x}{m^*}+\kappa\frac{\alpha_s^2k_x}{\sqrt{\alpha_s^2k_\perp^2+(\mu_nB_z)^2}}\equiv v_x
\end{align}
and
\begin{align}
    \partial_{k_y}E_\kappa(\bm{k}_\perp)
    &=\frac{k_y}{m^*}+\kappa\frac{\alpha_s^2k_y}{\sqrt{\alpha_s^2k_\perp^2+(\mu_nB_z)^2}}\equiv v_y,
\end{align}
where $v_{x,y}$ is the group velocity.
$s_\kappa^j(\bm{k}_\perp)$ can also be rewritten as 
\begin{align}
    s_\kappa^x(\bm{k}_\perp)=-\frac{\kappa}{2}\frac{\alpha_sk_y}{\sqrt{\alpha_s^2k_\perp^2+(\mu_nB_z)^2}},
\end{align}
\begin{align}
    s_\kappa^y(\bm{k}_\perp)=\frac{\kappa}{2}\frac{\alpha_sk_x}{\sqrt{\alpha_s^2k_\perp^2+(\mu_nB_z)^2}},
\end{align}
and 
\begin{align}
    s_\kappa^z(\bm{k}_\perp)=-\frac{\kappa}{2}\frac{\mu_nB_z}{\sqrt{\alpha_s^2k_\perp^2+(\mu_nB_z)^2}}.
\end{align}
Using them, we obtain the in-plane spin polarization
\begin{align}
    \delta S_x
    &=-\frac{(\nabla_\perp T)_y\alpha_s\tau}{2}
    \sum_{\kappa=\pm}
    \int\frac{d^2\bm{k}_\perp}{(2\pi)^2}
    \frac{k_y^2}{\sqrt{\alpha_s^2k_\perp^2+(\mu_nB)^2}}\cr
    &\quad\times
    \left(
    \frac{\kappa}{m^*}+\frac{\alpha_s^2}{\sqrt{\alpha_s^2k_\perp^2+(\mu_nB)^2}}
    \right)\frac{\partial f(E_\kappa(\bm{k}_\perp))}{\partial T},
\end{align}
where the term proportional to $(\nabla_\perp T )_x$ vanishes after the momentum integration.
Eventually, the Edelstein susceptibility reads
\begin{align}
    \chi_{xy}&=-\frac{\alpha_s\tau}{2}
    \sum_{\kappa=\pm}
    \int\frac{d^2\bm{k}_\perp}{(2\pi)^2}
    \frac{k_y^2}{\sqrt{\alpha_s^2k_\perp^2+(\mu_nB)^2}}\cr
    &\quad\times
    \left(
    \frac{\kappa}{m^*}+\frac{\alpha_s^2}{\sqrt{\alpha_s^2k_\perp^2+(\mu_nB)^2}}
    \right)
    \frac{\partial f(E_\kappa(\bm{k}_\perp))}{\partial T}
    \cr
    &=\frac{\alpha_s\tau}{8\pi}
    \sum_{\kappa=\pm}
    \int_0^{\infty} 
    \frac{k_\perp^3dk_\perp}{|\bm{d}(\bm{k}_\perp)|}
    \left(
    \frac{\kappa}{m^*}+\frac{\alpha_s^2}{|\bm{d}(\bm{k}_\perp)|}
    \right)\cr
    &\quad\times
    \left[-\frac{\partial f(E_\kappa(\bm{k}_\perp))}{\partial T}\right],
\end{align}
which is equivalent to the expression of $\chi_{xy}$ in the main text.
\end{document}